\documentclass[conf]{IEEEtran}
\usepackage{graphicx}
\usepackage{subcaption}
\usepackage{url}
\usepackage{amssymb,amsmath,bm}
\begin{document}
\title{LSTM Deep Neural Networks Postfiltering for\\Improving the Quality of Synthetic Voices}

\author{Marvin~Coto-Jim\'enez
        and John~Goddard-Close
\thanks{M. Coto-Jim\'enez, Department of Electrical Engineering, University of Costa Rica, San Jos\'e, Costa Rica and the Metropolitan Autonomous University, M\'exico D.F., M\'exico. 
e-mail: marvin.coto@ucr.ac.cr}
\thanks{John Goddard-Close, Department of Electrical Engineering, Metropolitan Autonomous University, M\'exico D.F., M\'exico. 
email:jgc@xanum.uam.mx}}
\maketitle

\begin{abstract}
Recent developments in speech synthesis have produced systems capable of outcome intelligible speech, but now researchers strive to create models that more accurately mimic human voices. One such development is the incorporation of multiple linguistic styles in various languages and accents.

HMM-based Speech Synthesis is of great interest to many researchers, due to its ability to produce sophisticated features with small footprint. Despite such progress, its quality has not yet reached the level of the predominant unit-selection approaches that choose and concatenate recordings of real speech. Recent efforts have been made in the direction of improving these systems.

In this paper we present the application of Long-Short Term Memory Deep Neural Networks as a Postfiltering step of HMM-based speech synthesis, in order to obtain closer spectral characteristics to those of natural speech. The results show how HMM-voices could be improved using this approach.
\end{abstract}


\section{Introduction}

Text-to-speech (TTS) synthesis is the technique of generating intelligible speech from a given text.  Applications of TTS have expanded from early supporting artifacts for the visually impaired, to in-car navigation systems, e-book readers, spoken dialog systems, communicative robots, singing speech synthesizers, and speech-to speech-translation systems~\cite{tokuda2013speech}. 

More recently, TTS systems have evolved from the sole production of intelligible voices to pursuit more sophisticated production of voices in multiple languages, with different styles and emotions~\cite{black2003unit}. Despite these trends, there are challenges, for example the overall quality of the voices. Researchers are striving to improve TTS systems by more closely mimicking natural human voices

The statistical methods for TTS, which arise in the late 1990s, have grown in popularity since then~\cite{yoshimuray1999simultaneous}, particularly those based on Hidden Markov Models (HMM), for their flexibility in changing speaker characteristics and low footprint, including capacities to produce average voices. HMM have been utilized extensively in speech recognition since about 30 years ago, as they provide a robust representation of the main events in which speech can be segmented~\cite{falaschi1989hidden}, with efficient parameter estimation algorithms. 

More than 30 reports of HMM-based Speech Synthesis implementations (also called Statistical Parametric Speech Synthesis) can be found for several languages around the world. For example~\cite{karabetsos2008hmm,pucher2010modeling,erro2010hmm,stan2011romanian,kuczmarski2010hmm,hanzlivcek2010czech,li2010hmm,vietnam2013,tamil2013,arab2015,jap2014,french2014}, are a few of the recent ones. Every implementation in a new language or its variants requires the adoption of HMM-related algorithms, incorporating their own linguistic specifications and making a series of decisions regarding the multiple definitions related to HMM, decision tress, and training conditions.

In this paper, we present our implementation of a statistical parametric speech synthesis system based on HMM with Long Short-Term Memory Postfilter Neural Networks for improving its spectral quality.

The rest of this paper is organized as follows: Section 2 provides some details of the HMM-based speech synthesis system; Section 3 presents the Long Short-Term Memory Neural Networks; and Section 4 describes the system and experiments carried out in order to test the Postfilter. Section 5 presents the results and analysis of objective evaluations, and the conclusions are in Section 6.

\section{Speech synthesis based on HMM}

HMM can be described from a Markov process, in which state transitions are given by a stochastic process. A second stochastic process models the emission of symbols when it comes to each state.

\begin{figure}
\begin{center}
\includegraphics[width=6cm]{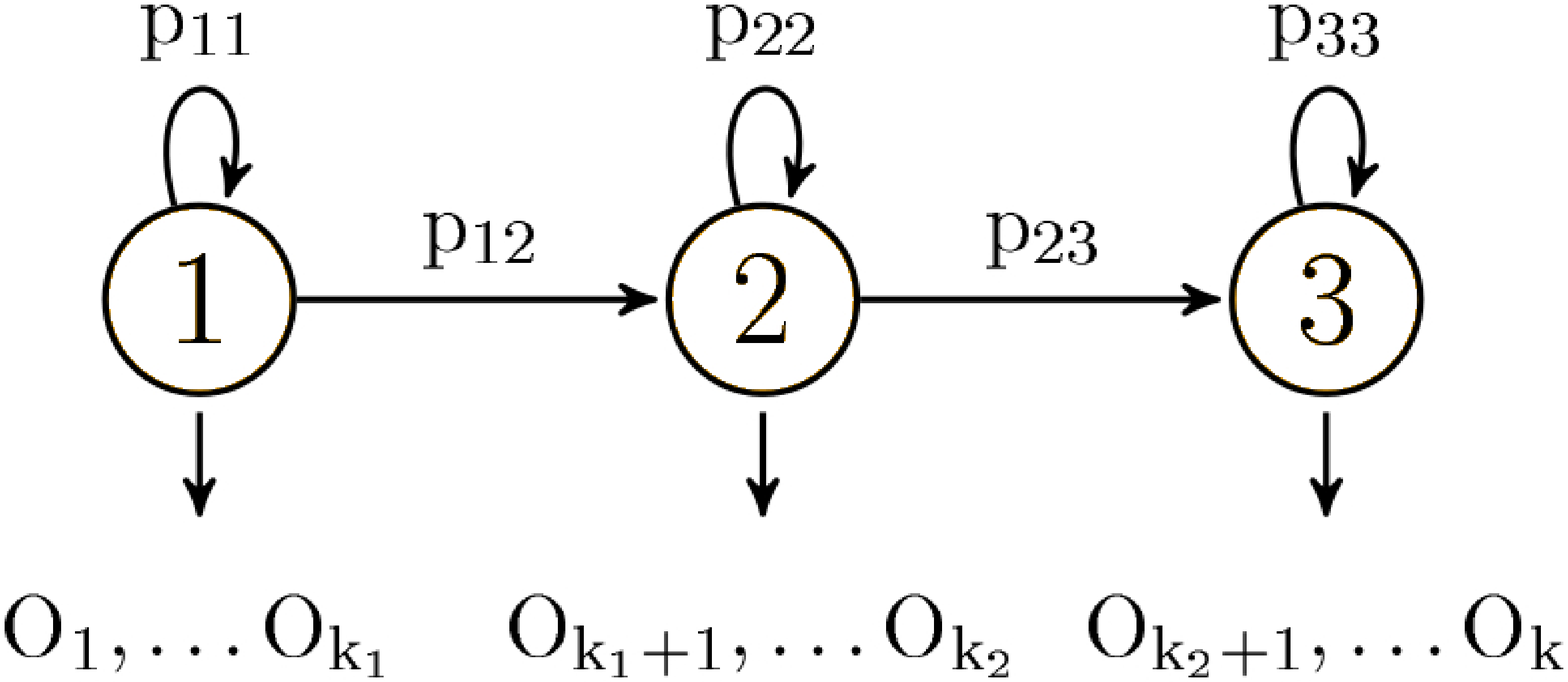}
\caption{Left to right example of an HMM with three states}
\label{hmm}
\end{center}
\end{figure}

In Figure~\ref{hmm}, a representation of a left to right HMM is shown, where there is a first state to the left from which transitions can occur to the same state or to the next on the right, but not in reverse direction. In this $p_{ij}$ represents the probability of transition from state $i$ to state $j$, and $O_{k}$ represents the observation emitted in state $k$.

In HMM-based Speech Synthesis, the speech waveforms can be reasonably reconstructed from a sequence of acoustic parameters learned and emitted as vectors from the HMM states~\cite{tokuda2013speech}. Typical implementation of this model includes vectors of observations with $f0$, MFCC and their delta and delta delta features for the adequate modeling of dynamic features of speech.

In order to improve the quality of the results, some researchers have recently experienced Postfiltering stages in which the parameters obtained with HTS voices have enhanced deep generative architectures~\cite{post1,post2,post3,post4}, for example DBM, RMB, BAM and recurrent neural networks.

In the next section, we present our proposal to incorporate Long Short-Term Memory Recurrent Neural Networks in the improvement of the quality of HMM-based speech synthesis.

\section{Long Short-Term Memory Recurrent Neural Networks}

mong the new algorithms to improve some tasks related to speech, such as speech recognition, groups of researchers have explored the use of Deep Neural Networks (DNN), with encouraging results. Deep learning, based on several kinds of neural networks with many hidden layers, have achieved great results in many machine learning and pattern recognition tasks. The disadvantage of using such networks is they cannot directly model the dependent nature of each sequence of parameters with the former, which is desirable to mimic the production of human speech. To solve this problem, it has been suggested to include RNN~\cite{lstm_4}~\cite{rnn_2} in which there is feedback from some of the neurons in the network, backwards or to themselves, forming a kind of memory that retains previous states.

An extended kind of RNN, which can store information over long or short time intervals, has been presented in~\cite{lstm2}, called Long Short-Term Memory (LSTM). LSTM was recently introduced to speech recognition, giving the lowest recorded error rates on the TIMIT database~\cite{lstm1}, among other successful applications of speech recognition~\cite{lstm3}.  The storage and use of long-term and short-term information is potentially significant for many applications, including speech processing, non-Markovian control, and music composition~\cite{lstm2}. 

In a RNN, output vector sequences $\mathbf{y}=\left(y_{1},y_{2},\dots,y_{T} \right)$ are computed from input vector sequences $\mathbf{x}=\left(x_{1},x_{2},\dots,x_{T}\right)$ and hidden vector sequences $\mathbf{h}=\left(h_{1},h_{2},\dots,h_{T} \right)$ iterating equations~\ref{eq:rnn1} and~\ref{eq:rnn2} from $1$ to $T$~\cite{lstm_4}:

\begin{equation}\label{eq:rnn1}
h_{t}=\mathcal{H}\left(\mathbf{W}_{xh}x_{t}+\mathbf{W}_{hh}h_{t-1}+b_{h} \right)
\end{equation}
\begin{equation}\label{eq:rnn2}
y_{y}=\mathbf{W}_{hy}h_{t}+b_{y}
\end{equation}

where $\mathbf{W}_{ij}$ is the weight matrix between layer $i$ and $j$, $b_{k}$ is the bias vector for layer $k$ and $\mathcal{H}$ is the activation function for hidden nodes, usually a sigmoid function $f:\mathbb{R}\rightarrow\mathbb{R},f(t)=\frac{1}{1+e^{-t}}$. 

Each cell in the hidden layers of a LSTM, has some extra gates to store values: an input gate, forget gate, output gate and cell activation, so values can be stored in the long or short term. These gates are implemented following the equations:

\begin{equation}
i_{t}=\sigma\left(\mathbf{W}_{xi}x_{t}+\mathbf{W}_{hi}h_{t-1}+\mathbf{W}_{ci}c_{t-1}+b_{i} \right)
\end{equation}
\begin{equation}
f_{t}=\sigma\left(\mathbf{W}_{xf}x_{t}+\mathbf{W}_{hf}h_{t-1}+\mathbf{W}_{cf}c_{t-1}+b_{f}\right)
\end{equation}
\begin{equation}
c_{t}=f_{t}c_{t-1}+i_{t}\tanh\left(\mathbf{W}_{xc}x_{t}+\mathbf{W}_{hc}h_{t-1}+b_{c} \right)
\end{equation}
\begin{equation}
o_{t}=\sigma\left(\mathbf{W}_{xo}x_{t}+\mathbf{W}_{ho}h_{t-1}+\mathbf{W}_{co}c_{t}+b_{o} \right)
\end{equation}
\begin{equation}
h_{t}=i_{t}\tanh\left(c_{t} \right)
\end{equation}
where $\sigma$ is the sigmoid function, $i$ is the input gate activation vector, $f$ the forget gate activation function, $o$ is the output gate activation function, and $c$ the cell activation function. $\mathbf{W}_{mn}$ are the weight matrices from each cell to gate vector. 

\section{Description of the system}

The resulting voices from the HTS system have notable differences with the original voices used in its production. Reducing the gap between natural and artificial voices can be learned directly from data~\cite{post1}. In our proposal, we use aligned utterances from natural and synthetic voices produced in the HTS system to establish correspondence between each frame. 

Given a sentence of natural speech and voice corresponding HTS, we extract a representation consisting of one coefficient for f0, one coefficient for energy, and 39 MFCC coefficients, using the system Ahocoder~\cite{aho}. The inputs to the LSTM network correspond to the MFCC parameters of each frame of the sentences produced with the HTS voice, while the output corresponds to the MFCC parameters of the natural voice of the same sentence. In this case, we have an exact correspondence between the vector, representing each phrase from HTS voice and natural voice by the alignment between both.

\begin{figure}[!h]
\begin{center}
\includegraphics[width=9cm]{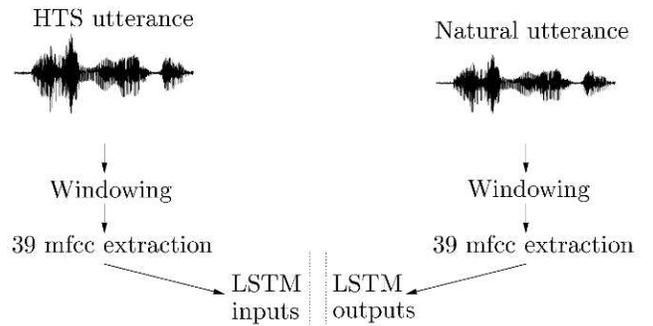}
\caption{Proposed system. HTS and Natural utterances are aligned frame by frame}
\label{fig:scheme}
\end{center}
\end{figure}

In this way, each LSTM Network attempts to solve the regression problem of transforming the values of the artificial speech and natural voice. This allows further improvement of the quality of new synthesized utterances with HTS, using this neural network as a subsequent step to approach these synthetic parameters to those of natural voice. Figure~\ref{fig:scheme} outlines the proposed system.

\subsection{Corpus description}
The CMU\_ARCTIC databases were constructed at the Language Technologies Institute at Carnegie Mellon University. They are phonetically balanced, with several US English speakers. It was designed for unit selection speech synthesis research.

The databases consist of around 1150 utterances selected from out-of-copyright texts from Project Gutenberg. The databases include US English male and female speakers. A detailed report on the structure and content of the database and the recording conditions is available in the Language Technologies Institute Tech Report CMU-LTI-03-177~\cite{cmu}. Four of the available voices were selected: BDL (male), CLB (female), RMS (male) and SLT (female).

\subsection{Experiments}

Each voice was parameterized, and the resulting set of vectors was divided into training, validation, and testing sets. The amount of data available for each voice are shown in Table ~\ref{table:data}. Despite all voices uttering the same phrases, the length differences are due to variations in the speech rate of each speaker.

\begin{table}[!h]
\begin{center}
\caption{Amount of data (vectors) available for each voice in the databases}
\label{table:data}
\begin{tabular}{p{1.3cm}p{1.5cm}p{1.5cm}p{1.5cm}p{1cm}}
Database & Total & Train & Validation & Test \\
\hline
  BDL   & 676554 & 473588  & 135311   &  67655    \\
  SLT   & 677970 & 474579  & 135594   &  67797    \\
  CLB   & 769161 &  538413 & 153832 &  76916    \\
  RMS   & 793067 & 555147  & 158613   & 79307     \\
\hline
\end{tabular}
\end{center}
\end{table}

The LSTM networks for each voice had three hidden layers, with 200, 160 and 200 units in each one respectively.

To determine the improvement in the quality of the synthetic voices, several objective measures were used. These measures have been applied in recent speech synthesis experiments and were found to be reliable in measuring the quality of synthesized voices~\cite{mcd1}~\cite{mcd2}:

\begin{itemize}
\item Mel Cepstral Distortion (MCD): Excluding silent phonemes, between two waveforms $v^{\mbox{targ}}$ and $v^{\mbox{ref}}$ it can be measured following equation~\ref{eq:mcd}~\cite{defmcd}

\begin{equation}\label{eq:mcd}
\mbox{MCD}\left(v^{\mbox{targ}},v^{\mbox{ref}}\right)=\frac{\alpha}{T}\sum_{t=0}^{T-1}\sqrt{\sum_{d=s}^{D}\left(v_{d}^{\mbox{targ}}(t)-v_{d}^{\mbox{ref}}(t) \right)^{2}}
\end{equation}

where $\alpha=\frac{10\sqrt{2}}{\ln 10}$, $T$ is the number of frames of each utterance, and $D$ the total number of parameters of each vector.

\item MFCC trajectory and spectrogram visualization: Simple observation of these elements by comparison permits to visualize quality in terms of similitude with those of the natural voice.

\end{itemize}

These measures were applied to the test set after being processed with the LSTM networks, and the results were compared with those of the HTS voices. The results and analysis are shown in the following section.

\section{Results and analysis}

For each synthesized voice produced with HTS and processed with LSTM networks, MCD results are shown in Table~\ref{table:mcd_results}. It can be seen how this parameter improved when all voices were processed with LSTM networks.
This shows the ability of these networks to learn the particular regression problem of each voice.

\begin{table}[!h]
\begin{center}
\caption{MCD between HTS and Natural Voices, and between LSTM Postfiltering and Natural Voice}
\label{table:mcd_results}
\begin{tabular}{p{2cm}p{2cm}p{2cm}}
Database & HTS to Natural & HTS to LSTM-PF  \\
\hline
  BDL   &   8.46                  &  7.98                  \\
  CLB   &   7.46                  &  6.87                  \\
  SLT   &   7.03                 & 6.65                  \\
  RMS   &   7.66                 &  7.60                 \\
\hline
\end{tabular}
\end{center}
\end{table}

\begin{figure}[!h]
\begin{center}
\includegraphics[width=0.4\textwidth]{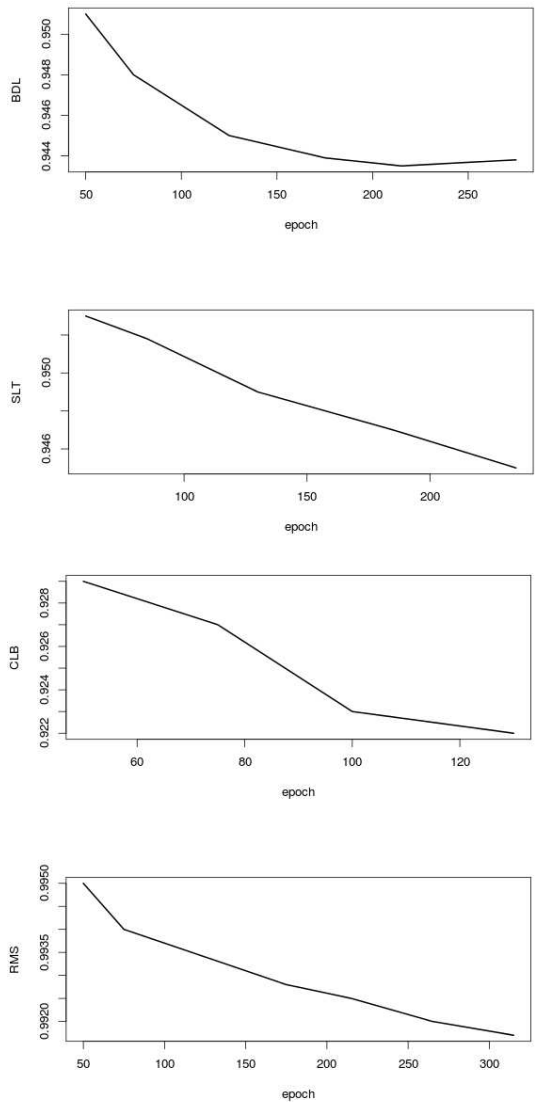}
\caption{Evolution of MCD improvement in LSTM Postfiltering during training epochs}
\label{fig:evolution}
\end{center}
\end{figure}

The best result of MCD improvement with the LSTM Postfiltering is CLB (7.9\%) and the least best was RMS(1\%). Figure~\ref{fig:evolution} shows how the MCD evolves with the training epochs for each voice. All HTS voices, except one, were improved by the LSTM Neural Network Postfilter in the MCD from the first 50 ephocs of training. 

\begin{figure}[!h]
\begin{center}
\includegraphics[width=0.5\textwidth]{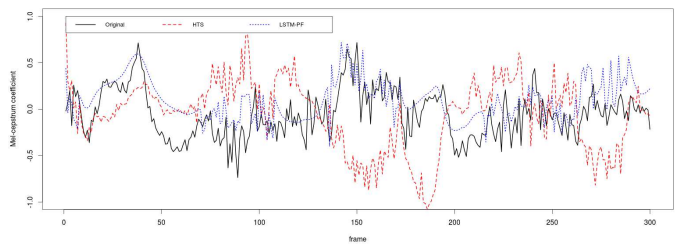}
\caption{Illustration of enhancing the 5th mel-cepstral coefficient trajectory by LSTM Postfiltering}
\label{fig:trajectory}
\end{center}
\end{figure}

\begin{figure}[!h]
    \begin{subfigure}[b]{0.4\textwidth}
        \includegraphics[width=\textwidth]{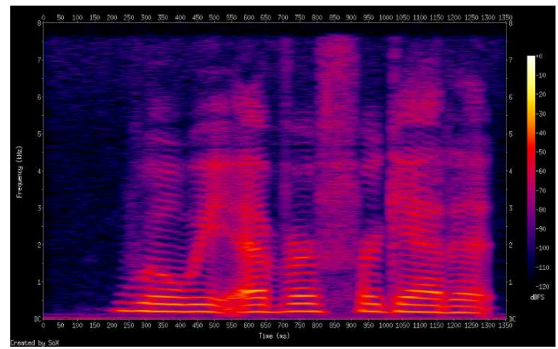}
        \caption{Original}
        \label{fig:hts}
    \end{subfigure}
    ~ 
    \begin{subfigure}[b]{0.4\textwidth}
        \includegraphics[width=\textwidth]{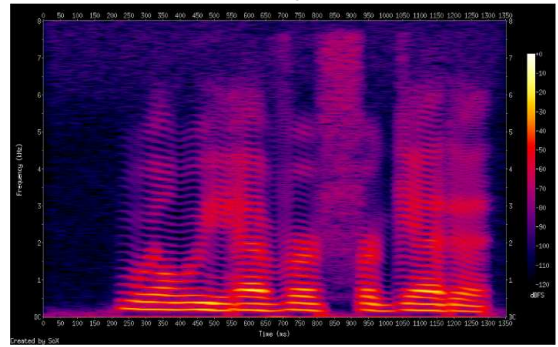}
        \caption{HTS}
        \label{fig:tiger}
    \end{subfigure}
    ~ 
    \begin{subfigure}[b]{0.4\textwidth}
        \includegraphics[width=\textwidth]{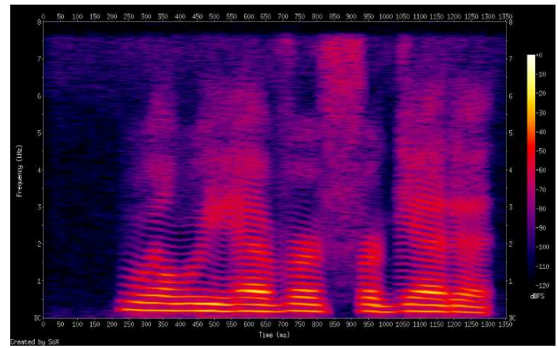}
        \caption{LSTM Postfiltering}
        \label{fig:lstm}
    \end{subfigure}
    \caption{Comparison of spectrograms}\label{fig:spectrograms}
\end{figure}

The differences in the amount of necessary epochs to reach convergence in each case are notable. This can be explaned by the difference in the MCD between HTS and natural voices. The gap between them is variable and the LSTM network requires more epochs to model the regression function between them.
An example of parameters generated by the HTS and the enhancement obtained by the LSTM Postfilter is shown in Figure~\ref{fig:trajectory}. It can be seen how the LSTM Postfilter fits the trajectory of the MFCC better than the HTS base system.
In Figure~\ref{fig:spectrograms} a comparison of three spectrograms of the utterance ``Will we ever forget it?'' for the HTS voice (a), Original (b) and LSTM Postfilter enhanced (c) is shown. The HTS spectrogram usually shows bands in higher frequencies not present in the natural voice, and the LSTM-Postfilter helps to smooth it, making it closer to the original voice spectrogram.

\section{Conclusions}

We have presented a new proposal to improve the quality of synthetic voices based on HMM with LSTM Postfiltering Networks. The LSTM have been able to learn directly from the data how to improve an artificial voice and make it mimic a more natural sound in its spectral characteristics.

We evaluated the proposed LSTM Postfilter in four voices, two masculine and two feminine, and the results show that all of them were improved in spectral features such as MCD measurement, spectrograms and mfcc trajectory generation.

The improvement of the HTS voices in MCD to the original voices were observed from the first training epochs of the LSTM neural network, but the convergence to a minimum distance took many more epochs. Due to the extensive amount of time required to train each epoch, further exploration should determine new network configurations or training conditions to reduce training time.

Future work will include the exploration of new representation of speech signals, hybrid neural networks and fundamental frequency enhancement with LSTM Postfilters.

\section{Acknowledgements}
  
This work was supported by the SEP and CONACyT under the Program SEP-CONACyT, CB-2012-01, No.182432, in Mexico, as well as the University of Costa Rica in Costa Rica. We also want to thank ELRA for supplying the original Emotional speech synthesis database.


\begin{thebibliography}{99}

\bibitem{tokuda2013speech}
Tokuda~K, Nankaku~Y, Toda~T, Zen~H, Yamagishi~J, and Oura~K (2013): Speech
  synthesis based on hidden markov models. In: Proceedings of the IEEE,
  101(5):1234--1252.

\bibitem{black2003unit}
Black~AW (2003): Unit selection and emotional speech. in \emph{INTERSPEECH}.

\bibitem{yoshimuray1999simultaneous}
Yoshimura~T, Tokuda~T, Masuko~T, Kobayashi~T and Kitamura~T (1999): Simultaneous modeling of spectrum, pitch and duration
  in HMM-based speech synthesis. In Proc. Eurospeech:2347--2350.

\bibitem{falaschi1989hidden}
Falaschi~A, Giustiniani~M, and Verola~M (1989): A Hidden Markov Model Approach to
  Speech Synthesis. In EUROSPEECH:2187--2190.

\bibitem{karabetsos2008hmm}
Karabetsos~S, Tsiakoulis~P, Chalamandaris~A, and Raptis~S (2008): HMM-based
  Speech Synthesis for the Greek Language. In Text, Speech and
  Dialogue. Springer, p 349--356.

\bibitem{pucher2010modeling}
Pucher~M, Schabus~D, Yamagishi~Y, Neubarth~F, and Strom~V (2010): Modeling and
  interpolation of Austrian German and Viennese Dialect in HMM-based Speech
  Synthesis. Speech Communication 52(2):164--179.

\bibitem{erro2010hmm}
Erro~D, Sainz~I, Luengo~I, Odriozola~I, S{\'a}nchez~J, Saratxaga~I,
  Navas~E, and Hern{\'a}ez~I (2010): HMM-based Speech Synthesis in Basque Language
  Using HTS. IN: Proceedings of the FALA.

\bibitem{stan2011romanian}
Stan~A, Yamagishi~Y, King~S, and Aylett~M (2011): The Romanian Speech Synthesis
  (RSS) Corpus: Building a High Quality HMM-based Speech Synthesis System Using
  a High Sampling Rate. In: Speech Communication, 53(3):442--450.

\bibitem{kuczmarski2010hmm}
Kuczmarski~T (2010): HMM-based Speech Synthesis Applied to Polish. Speech
  and Language Technology 12:13.

\bibitem{hanzlivcek2010czech}
Hanzl{\'\i}{\v{c}}ek~Z (2010):Czech HMM-based speech synthesis. In: Text,
  Speech and Dialogue. Springer, p 291--298.

\bibitem{li2010hmm}
Li~Y, Pan~S, and Tao~J (2010): HMM-based Speech Synthesis with a Flexible
  Mandarin Stress Adaptation Model. In: Proc. 10th ICSP2010 Proceedings,
  Beijing, p 625--628.

\bibitem{vietnam2013}
Phan ST, Vu TT, Duong CT and Luong MC (2013): A study in Vietnamese Statistical Parametric Speech Synthesis Based on HMM. International Journal, 2(1):p 1-6.


\bibitem{tamil2013}
Boothalingam~R, Sherlin Solomi~V, Gladston~AR, Christina~SL, Vijayalakshmi~P, Thangavelu~N, and Murthy HA (2013): Development and evaluation of unit selection and HMM-based speech synthesis systems for Tamil. In: National Conference on Communications (NCC), IEEE, p 1--5.

\bibitem{arab2015}
Khalil~KM and Adnan~C (2015): Implementation of speech synthesis based on HMM using PADAS database. In: 12th International Multi-Conference on Systems, Signals \& Devices (SSD). IEEE, p 1--6

\bibitem{jap2014}
Nakamura~K, Oura~K, Nankaku~Y, and Tokuda~K (2014): HMM-Based Singing Voice Synthesis and its Application to Japanese and English. In IEEE International Conference on Acoustics, Speech and Signal Processing (ICASSP), p 265--269.

\bibitem{french2014}
Roekhaut~S, Brognaux~S, Beaufort~R, and Dutoit~T (2014): Elite-HTS: a NLP tool for French HMM-based speech synthesis. In: Interspeech, p 2136--2137.

\bibitem{post1}
Chen~LH, Raitio~T, Valentini-Botinhao~C, Ling~ZH and Yamagishi~J (2015): A deep generative architecture for postfiltering in statistical parametric speech synthesis. IEEE/ACM Transactions on Audio, Speech and Language Processing (TASLP), 23(11):2003--2014.

\bibitem{post2}
Takamichi~S, Toda~T, Neubig~G, Sakti~S and Nakamura~S (2014): A postfilter to modify the modulation spectrum in HMM-based speech synthesis. In: IEEE International Conference on Acoustics, Speech and Signal Processing (ICASSP), p 290-294.

\bibitem{post3}
Takamichi~S, Toda~T, Black~AW and Nakamura~S (2014): Modified post-filter to recover modulation spectrum for HMM-based speech synthesis. In IEEE Global Conference on Signal and Information Processing (GlobalSIP), p 547--551.

\bibitem{post4}
Prasanna Kumar~M and Black~AW (2016): Recurrent Neural Network Postfilters for Statistical Parametric Speech Synthesis. arXiv preprint arXiv:1601.07215.

\bibitem{lstm_4} 
Fan~Y, Qian~Y, Xie~FL and Soong~FK (2014): TTS synthesis with bidirectional LSTM based recurrent neural networks. In Interspeech, p 1964--1968.

\bibitem{rnn_2} 
Zen~H and Sak~H (2015): Unidirectional long short-term memory recurrent neural network with recurrent output layer for low-latency speech synthesis. In: IEEE International Conference on Acoustics, Speech and Signal Processing (ICASSP), p 4470--4474.

\bibitem{lstm2} 
Hochreiter~S and Schmidhuber~J (1997): Long short-term memory. Neural computation 9(8): 1735--1780.

\bibitem{lstm1} 
Graves~Alan, Jaitly~N, Mohamed~A (2013): Hybrid speech recognition with deep bidirectional LSTM. In: IEEE Workshop on Automatic Speech Recognition and Understanding (ASRU).

\bibitem{lstm3}
Graves~A, Fern\'andez~S and Schmidhuber~J. (2005): Bidirectional LSTM networks for improved phoneme classification and recognition. Artificial Neural Networks: Formal Models and Their Applications--ICANN. Springer Berlin Heidelberg, p 799--804.

\bibitem{aho}
Erro~D, Sainz~I, Navas~E, Hernaez~I (2011): Improved HNM-based Vocoder for Statistical Synthesizers. InterSpeech, p 1809--1812. 

\bibitem{cmu}
Kominek~J and Black~AW (2004): The CMU Arctic speech databases. Fifth ISCA Workshop on Speech Synthesis.

\bibitem{mcd1}
Zen~H, Senior~A, and Schuster~M (2013): Statistical parametric speech synthesis using deep neural networks. In: IEEE International Conference on Acoustics, Speech and Signal Processing (ICASSP).

\bibitem{mcd2}
Zen~H and Senior~A (2014): Deep mixture density networks for acoustic modeling in statistical parametric speech synthesis. In: IEEE International Conference on Acoustics, Speech and Signal Processing (ICASSP).

\bibitem{defmcd}
Kominek~J, Schultz~T and Black~AW (2008): Synthesizer voice quality of new languages calibrated with mean mel cepstral distortion. SLTU.

\end{thebibliography}
\end{document}